\documentclass[aps,prd,amssymb,showpacs,twocolumn]{revtex4}
\usepackage{epsfig}
\usepackage[latin1]{inputenc}
\usepackage{amsmath}

\begin{document}
\title{Dealing with delicate issues in waveforms calculations}

\author{Luis Lehner${}^1$\thanks{
email: lehner@phys.lsu.edu}
and
Osvaldo M. Moreschi${}^2$\thanks{Member of CONICET.}
\thanks{ email: moreschi@fis.uncor.edu} \\
\vspace{3mm} 
${}^1$\small Department of Physics \& Astronomy, LSU, \\
\small 16224 Baton Rouge, LA 70810.  \\
\vspace{3mm} 
${}^2$\small FaMAF, Universidad Nacional de C\'{o}rdoba\\
\small Ciudad Universitaria,
(5000) C\'{o}rdoba, Argentina. 
}

\begin{abstract}
We revisit the calculation of gravitational
radiation through the use of Weyl scalars. We 
point out several possible problems arising from gauge and
tetrad ambiguities and ways to address them. Our analysis
indicates how, relatively simple corrections can be
introduced to remove these ambiguities.
\end{abstract}

\maketitle



\section{Introduction}
The definition of gravitational radiation involves a fair amount
of work so as to distinguish variations in the metric that are
purely coordinate effects from true signals propagating on a background
which itself must be well established. 
At finite (but sufficiently far) distances from an isolated
source several approaches exist
to do so provided a suitable background can be identified\cite{moncrief,wheeler,zerilli,teukolsky,abrahamsprice}.
For sufficiently generic cases the unambiguous identification of the gravitational
radiation can only be done at future null infinity (${\cal I}^+$). This is possible
since when studying an isolated system, future null infinity is the
asymptotic region of an asymptotically flat spacetime. There the metric is
exactly flat and disturbances around it can be associated with outgoing
gravitational waves subject to singling out a preferred frame. This issue
arises as there is no unique flat asymptotic metric. In fact there are as many 
of them as supertranslations are in the BMS group; the asymptotic
symmetry group. 
Nevertheless there is a well defined notion of gravitational
radiation which exploits 
the algebraic notion of asymptotic flatness\cite{Moreschi87} which in turn  
implies the existence of the gravitational radiation fields 
$\Psi_4^0$ and $\Psi_3^0$. The existence of these fields can also
be deduced from stronger conditions, like peeling\cite{newmanpenrose62,newmanunti}.
Then, after a careful choice of coordinates, radiation can be unambiguously defined.

Unfortunately, numerical applications dealing with black hole spacetimes can not
yet reach ${\cal I}^+$ unless Cauchy-characteristic extraction 
is employed\cite{extractionI,extractionII} for the extraction procedure or
Cauchy-characteristic matching\cite{matching} 
or the conformal formulations\cite{friedrich} are fully realized 
to model the whole spacetime.  Without adopting any of these options a commonly employed approach
for obtaining gravitational waves relies on applying the infrastructure developed for ${\cal I}^+$ at finite
distances. Here, several compromises are made as natural
quantities defined at ${\cal I}^+$ need to be translated somehow to finite
distances where they need not be well defined.

The purpose of this note is to point out difficulties that likely arise in
this approach and a route to address them.

\section{Review of the approach}
The calculation of gravitational radiation at ${\cal I}^+$ is based on the
foundational works of Bondi\cite{bondi} and Sachs\cite{sachs} as an approximation expansion at infinite
distances and by the work of Penrose\cite{penrose} as a geometrical construct at the boundary
of the physical spacetime. Under reasonable assumptions one can show that
the spacetime is asymptotically flat and 
time dependent perturbations close to the boundary are either due to 
gravitational waves or coordinate effects. In order to disentangle coordinate with physical effects,
care must be taken in dealing with the asymptotic symmetry group and fixing a convenient
frame with respect to which inertial observers can define radiation.

Our goal here is not to revisit this approach (of which descriptions
have been presented 
in e.g. \cite{newmantod,newmanunti,newmanunti2,tamburino,winicour,moreschi86}), but simply to state
the crucial ingredients which are missing when the strategy is employed at 
finite distances. In the next sections we will carry out an analysis of 
what corrections are required when key assumptions described below are missing.

These ingredients are,
\begin{itemize}
\item Peeling is assumed, this means that suitable Weyl curvature components
behave asymptotically in a well defined manner. 
\item Outgoing null hypersurfaces, parameterized by $u$ intersect ${\cal I}^+$
defining a sequence of $S^2$ surfaces. 
\item Each of these surfaces is conformal to a unit sphere metric; off this surface (into
the spacetime) the departure from it is of lower order. Namely the angular
metric in a neighborhood of ${\cal I}^+$ can be expressed, 
as 
$g_{AB} = r^2 h_{AB} =  r^2 (q_{AB}/V^2 + c_{AB}/r + O(r^{-2})) $; 
with
$V$ a conformal factor, $q_{AB}$ a unit sphere metric, and $r$ a suitably
defined radial distance.
\item A null-tetrad $\{\ell^a,n^a,m^a,\bar m^a\}$ satisfying
$\ell^a n_a = - m^a \bar m_a = 1$ (with all other products being $0$).  The 
spacetime metric can be expressed as $g_{ab} = 2 l_{(a} n_{b)} - 2 m_{(a} \bar m_{b)}$,
and suitable projections of the Weyl tensor can be defined. 
\item Using standard\cite{Pirani64,Geroch73} 
conventions for the Riemann tensor and spinor dyad,
particularly useful scalars obtained from them 
are,
\begin{eqnarray} 
\Psi_4 &=&  C_{abcd} n^a \bar m^b n^c \bar m^d\\
\Psi_3 &=&  C_{abcd} \ell^a n^b \bar m^c \bar n^d\\
\Psi_2 &=&  C_{abcd} \ell^a m^b \bar m^c n^d  \\
\sigma &=& m^a m^b \nabla_a l_b.
\end{eqnarray}
\item A suitable (Bondi type) expansion in terms of $1/r$ coupled to 
coordinates chosen such that $V=1$, $g_{ur}^0=1$ and $g_{uA}^0=0$ ($x^A$ labeling
angular coordinates at $u=const$, $r \rightarrow \infty $) 
gives rise to several important relations~\cite{newmanunti,Moreschi87}. 
In particular,
\begin{eqnarray}
\Psi_4^0 &=& -\ddot{ \bar \sigma }^0 \label{eqp4sigma} \\
\Psi_3^0 &=& - \eth \dot{\bar\sigma}^0  \\
\dot \Psi_2^0 &=&  \eth \Psi_3^0 + \sigma^0 \Psi_4^0 ;
\end{eqnarray}
where the supra-index ``${}^0$'' indicates leading order in an expansion in the radial coordinate
and $\eth$ is the edth operator\cite{Geroch73} of the
unit sphere.
\end{itemize}
With this structure at hand, the following results are obtained.
Given any section $S$ at future null infinity, the Bondi momentum
is given by
\begin{equation}
P^a =  - \frac{1}{4 \pi} \int_S \hat l^{a} (\Psi_2^0 + \sigma^0 \dot {\bar \sigma}^0) dS^2 ,
\end{equation}
where
$\hat l^a =(1,\sin(\theta) \cos(\phi), \sin(\theta) \sin(\phi), \cos(\theta) ) $ when expressed in standard 
angular coordinates $(\theta,\phi)$.
The so called Bondi mass
$M$ is the timelike component of this vector; namely
\begin{equation}
M =  - \frac{1}{4 \pi} \int_S (\Psi_2^0 + \sigma^0 \dot {\bar \sigma}^0) dS^2 .
\end{equation}
It is interesting to note that the Bondi momentum can also be
expressed in the terms of the Psi supermomentum\cite{moreschi88}
$\Psi \equiv \Psi_2^0 + \sigma^0 \dot {\bar \sigma}^0 
+ \eth^2 \bar\sigma^0$ by
\begin{equation}
P^a =  - \frac{1}{4 \pi} \int_S \hat l^{a} \Psi dS^2 ;
\end{equation}
since it has a couple of
useful properties; namely, it is real,
$\bar\Psi = \Psi$, and its time derivative is simply
\begin{equation}
\dot \Psi = \dot \sigma^0 \dot {\bar \sigma}^0.
\end{equation}
Therefore, the time variation of the Bondi momentum,
in terms of time Bondi coordinate, is just
given by
\begin{equation}
\dot P^a =  - \frac{1}{4 \pi } \int_S \hat l^{a} \dot \sigma^0 \dot {\bar \sigma}^0 dS^2 \label{radiate} .
\end{equation}

Notice that due to
relation (\ref{eqp4sigma}) one might choose to replace 
any appearance of $\dot \sigma^0$ by the time integral of $\Psi_4^0$.
This is commonly done in numerical 
simulations \cite{sources,loustolazzarus,pretorius,loustowaves,nasawaves,bruegman,bosonwavesI,loustoJ,Pfeiffer:2007yz,Koppitz:2007ev};
however one should notice that (\ref{eqp4sigma}) is only valid
in terms of a Bondi tetrad and coordinate system.

\section{Connection with finite distances}
The above expressions provide a well defined infrastructure with which gravitational radiation
can be defined. However, to obtain them one has introduced key ingredients, otherwise the
expressions would result significantly more involved. In particular without choosing 
the conformal frame
and/or radial distance such that the leading part of the angular metric is indeed that of the
unit sphere, several changes arise even when all other properties are satisfied. 
For instance, if the angular metric were given by 
$g_{AB} = r^2 ( q_{AB} / V(u,\theta, \phi)^2 + O(r^{-1})
)$ 
then, instead of (\ref{eqp4sigma}) 
one would have
\begin{equation}
\begin{split}
\Psi_4^0 &=
-\ddot{ \bar \sigma }^0
-\frac{1}{V}\bar\eth^2 \dot V
+\frac{2}{V^2}\bar\eth \dot V \bar\eth V
-\frac{2}{V^3} \dot V (\bar\eth V)^2 \\
&
+\frac{1}{V^2}\dot V \bar\eth^2 V
-\frac{3}{V^2} \dot V^2 \bar\sigma^0
+\frac{1}{V}\ddot V \bar\sigma^0
+\frac{3}{V}\dot V \dot{\bar\sigma}^0 \\
&=
-\ddot{ \bar \sigma }^0
-\bar\eth^2 \frac{\dot V}{V}
-\frac{3}{V^2} \dot V^2 \bar\sigma^0
+\frac{1}{V}\ddot V \bar\sigma^0
+\frac{3}{V}\dot V \dot{\bar\sigma}^0 .
\end{split}
\end{equation}
Therefore, if we call $\tilde \Psi_4^0$
the inertial (Bondi) radiation field one would
have the relation
\begin{equation}\label{eq:tpsi4bondi2}
\tilde \Psi_4^0 
= \frac{1}{V^3} 
\left(
-\ddot{ \bar \sigma }^0
-\bar\eth^2 \frac{\dot V}{V}
-\frac{3}{V^2} \dot V^2 \bar\sigma^0
+\frac{1}{V}\ddot V \bar\sigma^0
+\frac{3}{V}\dot V \dot{\bar\sigma}^0 
\right) .
\end{equation}
where the factor $V^{-3}$ results from one of the corrections
which we discuss in this work.
Notice that this is not a perverse circumstance, rather it can be
generically expected as surfaces at $u=\text{const},r=\text{const}$ 
define a 2-sphere
whose metric is always conformally related to the unit sphere.

Certainly, in principle coordinate conditions can be adopted so that at the extraction sphere
the simplifying conditions hold, these conditions need not be those that  are convenient for the
numerical implementations. Indeed, coordinate conditions are exploited in a non-trivial manner
to aid in the numerical simulation\cite{aeigauge,nasagauge,pretoriusgauge}. Therefore,
one is then left with having to consider how to correctly identify suitable coordinates and/or a frame
to extract the desired quantity. Here, one of two options can be adopted. One approach 
is motivated from a perturbative point of view and relies in extracting a suitable Kinnersley
tetrad\cite{Kinnersley} from the numerically generated spacetime\cite{loustoKinners,nerozziKinners,burkoKinners}.
The so called Kinnersley tetrads are null tetrads that 
were constructed for the study of type D vacuum spacetimes;
but one can in general choose them so that
to leading order they conform to a Bondi tetrad.
The basic idea is to use a null tetrad adapted to the asymptotic
structure so that in the limit coincides with a Bondi tetrad.
Armed with this tetrad
the computation of the radiative properties of the spacetime is then well defined.
The delicate point however stems from the fact that this approach requires a 
suitable numerical identification of a Bondi frame where errors can arise.
The influence of these errors 
in a related context has been recently discussed in\cite{pazos}.

A different approach
is to consider the asymptotic structure of asymptotically flat spacetimes and define the analog
quantities at finite distances. This approach, which we adopt here, does not rely on obtaining
background quantities correctly, rather one needs only identify a suitable Bondi frame which will
essentially be unambiguous as long as the extraction worldtube is {\it sufficiently far}.
We will not concern ourselves with how far must the extraction worldtube be, rather we assume one is able to
place it sufficiently far for the peeling property, in the extracted quantities, of the Weyl scalars to be observed.
Under this assumption we analyze what consequences, and most importantly, modifications one must take into
account when calculating physical observables with the coordinates naturally induced
on the extraction worldtube in a numerical simulation.

To this end, it will be important to calculate the induced metric on the extraction worldtube
and evaluate different quantities which play a key role. We begin by following precisely the
same  procedure commonly employed in numerical simulations (see for 
instance\cite{sources,lazarus,bruegman}).  First a Cartesian extraction worldtube 
is defined by $x^2 + y^2 + z^2 = R^2$.
Observers at this worldtube are defined by their trajectories
on the worldtube given by $(t,x=\text{const.},y=\text{const.},z=\text{const.})$.
Then,
$\Psi_4$ is calculated on the hypersurface $\Sigma_t$  (defined by $t=\text{const.}$) by suitably
 adopting a tetrad frame.
 This tetrad is chosen in a
straightforward manner by adopting three orthonormal spatial vectors 
$\{ r^a,\theta^a, \phi^a \}$ on $\Sigma_t$  
(which are the analogs to the radial and tangents vectors to the sphere) 
and combine them with the unit normal vector $N^a$ to the hypersurface.
Once $\Psi_4$ is obtained, an interpolation on to the
extraction sphere is carried out. 
At this point it is useful to recall that the extraction must be done along
null rays, thus any comparison (or refinements of the extracted quantities\footnote{For instance
by employing the extracted values at different radii to determine the leading behavior
of $\Psi_4$ by fitting a suitable polynomial in $R^{-1}$.}) at
different worldtubes should contemplate suitable time offsets.
This offset reflects the arrival time of a null signal from a given internal worldtube
to the next.

As mentioned, this procedure, need not give rise to a structure fully compatible
with those mentioned in section I. In particular the induced angular sphere
metric need not be sufficiently close to that of the unit sphere, 
inertial observers coordinates might `shift' in time around the worldtube and their associated times 
might tick at different rates. At finite distances these issues are even more relevant
since the waves themselves will influence the geometry of the extraction worldtube as
they propagate across it. In the next section we discuss the corrections required
and will illustrate their application in  
section V. As we will see,  even in linearized problems these issues can play a non-trivial role.

A short summary of what these corrections entail is (in addition to the standard
first-step to calculate $\Psi_4$) is,

\begin{itemize}

\item Assuming that the extraction world tube has been chosen
far enough\footnote{By calculating  $\Psi_4$
at different extraction radii, this radial behavior can be checked}, 
calculate the leading order behavior $\Psi_4^0$
from  $\Psi_4 = \Psi_4^0 r^{-1} + O(r^{-2})$;
where $r$ is an appropriate asymptotic radial coordinate;
such that $r=R$ at the extraction world tube.

\item In a similar way, obtain the leading order behavior
of the induced metric  $g_{\Gamma}$ on the sphere at the extraction worldtube
defined by $t=\text{const.}$, $r =R$,
from 
$g_{\Gamma}= -q R^2 V^{-2} + O(r^{1})$;
where $q$ is the unit sphere metric.

\item Obtain the leading order behavior $g_{ur}^0$
from the expansion $g_{ur} = g_{ur}^0 + O(r^{-1})$;
 which measures the time
observers of $\Psi_4$ have.

\item Obtain the leading order behavior $g_{uA}^0$
from the expansion $g_{uA} = g_{uA}^0 + O(r^{-1})$; which measures the coordinate shift
in time observers of $\Psi_4$ have around the worldtube.

\item Bondi's radiation $\tilde \Psi_4^0$ is then obtained by relatively simple correction factors
from the knowledge of $\{g_{ur}^0,g_{uA}^0,V \}$. In the particular case where $g_{uA}^0\simeq 0$
the expression is
\begin{equation}
\tilde \Psi_4^0 = \frac{1}{(g_{ur}^0)^2 V^3} \Psi_4^0
\end{equation}
\end{itemize}

\section{Asymptotic structure of asymptotically flat spacetimes}\label{sec:asymptotic}
In this section we examine which corrections must be taken into account to
remove the ambiguities described previously. To this end, we will
examine how the relevant quantities transform among different coordinate systems
compatible with the asymptotic structure. Then, by adopting one of these systems to
be a Bondi one, we will be able to extract the correcting terms. For the sake of
simplicity in the derivation we adopt the spinor-calculus approach, though the results
obtained are completely independent of this technique.

We begin by determining key relations between different frames, to do so
it is convenient to have at hand some basic equations of the asymptotic structure.

\subsection{Basic variables}
The asymptotic geometry can be expressed in terms of a complex null tetrad
$\left( \ell ^{a},m^{a},\overline{m}^{a},n^{a}\right) $ with the properties:
\begin{equation}\label{eq:produc}
g_{ab}\;\ell ^{a}\;n^{b}=-g_{ab}\;m^{a}\;\overline{m}^{b}=1
\end{equation}
and all other possible scalar products being zero, the metric can be
expressed by
\begin{equation}
g_{ab}=\ell _{a}\;n_{b}+n_{a}\;\ell _{b}-m_{a}\;\overline{m}_{b}-\overline{m}%
_{a}\;m_{b}.
\end{equation}
Such a null tetrad is 
easily related to a dyad of spinors $(o^A, \iota^A)$.
The relation of the null tetrad with a spinor dyad is given by
$\ell^a \Longleftrightarrow o^A o^{A'}$ , 
$m^a \Longleftrightarrow o^A \iota^{A'}$ , 
$\bar m^a \Longleftrightarrow \iota^A o^{A'}$  
and
$n^a \Longleftrightarrow \iota^A \iota^{A'}$.
Thus, determining how the tetrad trasnforms one infers the
transformation of $(o^A,\iota^A )$. In our discussion, to make a direct contact with
the standard literature on the subject we 
also adopt a null polar coordinate system $(x^{0},x^{1},x^{2},x^{3})=\left( u,r,%
(\zeta +\overline{\zeta}),\frac{1}{i}(\zeta -\overline{\zeta})\right)$.  However,
as opposed to the standard discussions we will not make use of the available
coordinate freedom to simplify the treatment at the onset. Rather, we will adopt coordinates
consistent with those employed in the extraction procedure at finite
distances and deal with the consequences of this choice.

In most numerical applications, one normally works with finite
size regions; and would like to estimate the asymptotic fields
in terms of null tetrads based on a choice of coordinate system.
At the extraction worldtube, one assumes that a coordinate system
$(t,r,\zeta, \bar\zeta)$ can be constructed. Let $\Gamma$, be the timelike
surface defined by $(r=R=\text{const.})$.
On $\Gamma$, a null tetrad $\ell,n,m,\bar m$ can be defined in the following way.
First, define the null function $u$ such that on $\Gamma$
one has $u=t$; and $\ell=du$ everywhere. The function $u$ is chosen
such that the future directed vector $\ell$ points outwards with respect to the 2-surfaces
$(t=\text{const.},r=R)$; which are, topologically,
two dimensional spheres.

The complex vectors $m$ and $\bar m$ are defined to be
tangent to the spheres $(t=\text{const.},r=\text{const.})$. 
Furthermore, one can choose $m$ to be proportional
to $\frac{\partial}{\partial \zeta}$; and
$\bar m$ to be proportional
to $\frac{\partial}{\partial \bar\zeta}$
in the asymptotic region for large $r$.

Then, by requiring the standard normalization conditions
given in (\ref{eq:produc}) one settles the remaining
null vector $n^a$.

The coordinate system  $(u,r,\zeta, \bar\zeta)$ is thus straightforwardly
related to the one commonly employed in numerical efforts $(t,R,\theta,\phi)$ though
we will continue our discussion with the former as it is the one employed
in the standard literature on the subject. The conclusion however, will be independent
of this choice.

Keeping $(\zeta=\text{const.})$ and $(\bar\zeta=\text{const.})$
on the null hypersurface $u=\text{const.}$,
and increasing $r$ one moves along a null
direction. Since $\ell$ is contained on 
the hypersurface $u=\text{const.}$, one deduces that
$\ell$ is proportional to $\frac{\partial}{\partial r}$.
Then, one can write
\begin{equation}
\left(\ell^a\right) = 
\left(\frac{1}{g_{ur}}\frac{\partial}{\partial r}
\right)^a ;
\end{equation}
The appearance
of $g_{ur}$ in the denominator is the first correction to 
be accounted for due to the structure induced at $\Gamma$
not necessarily conforming to a Bondi frame.
The intersection of the null hypersurface $u=\text{const.}$
with future null infinity defines a two dimensional surface
denoted by $S$.
A natural null tetrad is then completed with,
\begin{equation}
\ell _{a}=\left( du\right)_{a} 
\label{uno}
\end{equation}
\begin{equation}
m^{a}=\xi ^{i}\left( \frac{\partial }{\partial x^{i}}\right) ^{a} 
\label{eq:vecm}
\end{equation}
\begin{equation}
\overline{m}^{a}=\overline{\xi}^{i}\left( \frac{\partial }{\partial x^{i}}\right) ^{a} 
\label{tres}
\end{equation}
\begin{equation}\label{eq:vecn}
n^{a}=\,\left(\frac{\partial}{\partial \,u} \right)^{a}
+ \,U\,\left( \frac{\partial }{\partial \,r}\right)^{a}
+ X^{i}\,\left(\frac{\partial }{\partial \,x^{i}}\right)^{a} 
\end{equation}
with $i=2,3$  and components $\xi^{i}$, $U$ and $X^{i}$ are:
\begin{equation}
\xi ^{2}=\frac{\xi _{0}^{2}}{r} + O\left(\frac{1}{r^2}\right),
\qquad \xi ^{3}=\frac{\xi_{0}^{3}}{r} + O\left(\frac{1}{r^2}\right) , 
\end{equation}
with
\begin{equation}\label{eq:xileading}
\xi _{0}^{2}=\sqrt{2}P_{0}\;V,\qquad \xi _{0}^{3}=-i\xi _{0}^{2},
\end{equation}
where $V=V(u,\zeta,\bar\zeta)$ and
the square of $P_0 = \frac{(1+\zeta \bar\zeta)}{2}$ is the conformal factor 
of the unit sphere;
\begin{equation}
U=rU_{00}+U_{0}+\frac{U_{1}}{r} +  O\left(\frac{1}{r^2}\right), 
\end{equation}
where
\begin{equation}
U_{00}=\frac{\dot{V}}{V},\quad U_{0}=-\frac{1}{2}K_{V},\quad U_{1}=-\frac{%
\Psi _{2}^{0}+\overline{\Psi}_{2}^{0}}{2}, 
\end{equation}
where $K_V$ is the Gaussian curvature,
 given by
\begin{equation}
\begin{split}\label{eq:kurv1}
K_{V}&= 2 V ~\overline{\eth }\eth\, V - 2 \eth V~%
\bar{\eth} V + V^{2} \\
&= V \nabla^2 V - \nabla^i V \, \nabla_i V + V^2 \\
&= V^2 \nabla^2 \ln V + V^2 
,
\end{split}
\end{equation}
of the 2-metric
\begin{equation}\label{eq:desphere}
dS^{2}=\frac{1}{V^2 \, P_0^{2}}\;d\zeta \;d\bar{\zeta} ;
\end{equation}
where the regular conformal metric restricted to $S$ is precisely
$\tilde g \mid_{S} = - dS^2$,
$\nabla^2 = \nabla^i \nabla_i$ is the Laplacian operator of the unit sphere
metric and $\nabla_i$ its covariant derivative.
Let us emphasize that $V=1$ makes $dS^2$ in (\ref{eq:desphere}) the
unit sphere metric.
Finally, the other components of the vector $n^a$ have the asymptotic form
\begin{equation}
 X^{2}= X_2^{0} + O\left(\frac{1}{r^2}\right) ,
\qquad X^{3}= X_3^{0} + O\left(\frac{1}{r^2}\right)  .
\end{equation}

\subsection{Asymptotic gauge freedom, restricted case}
At this point, we find it convenient to restrict to a simplified
case before describing the general one. We will assume here
that $X_2^{0}=X_3^{0}=0$ which, in an alternative view implies that
angular coordinates at $\Gamma$ do not shift in time as $g_{uA}^0=0$.
Our discussion will center around the allowed transformations between
the induced coordinates at the worldtube $(u,r,\zeta,\bar \zeta)$
and the Bondi system $(\tilde u,\tilde r,\tilde \zeta,\bar {\tilde \zeta})$.

\subsubsection{Coordinate and tetrad transformations I}
Let us consider the main gauge freedom admitted in our calculation 
which is of the form
\begin{align} 
\tilde u &= \alpha(u,\zeta,\bar\zeta) + 
 \frac{\tilde u_1(u,\zeta,\bar\zeta)}{r}  \label{eq:tildeu}
+ O\left(\frac{1}{r^2}\right) ,\\
\tilde r &= \frac{r}{w(u,\zeta,\bar\zeta)}  \label{eq:tilder}
+ O\left(r^0\right)  ,\\
\tilde \zeta & = \zeta + O\left(\frac{1}{r}\right) .\label{eq:tildezeta}
\end{align}
with $\dot \alpha >0$.
The possible further transformation of the coordinates of the sphere
$(\zeta,\bar\zeta)$ into itself is not needed at this point.

The condition $g^{\tilde u \tilde r}=1$ in a Bondi system imposes the relation
\begin{equation}
w = \frac{\dot\alpha}{g_{ur}^0} .
\end{equation}

This asymptotic coordinate transformation is associated to a corresponding
null tetrad transformation; which to leading orders is given by
\begin{equation}
\begin{split}
\tilde \ell & = d\tilde u = \dot\alpha \, du
+ \alpha_\zeta  \, d\zeta + \alpha_{\bar\zeta}  \, d\bar\zeta 
+O\left(\frac{1}{r}\right) \\
&=  \dot\alpha  \, \ell
- \frac{\eth_{V} \alpha}{r}  \, \bar m 
- \frac{\bar\eth_V \alpha}{r}  \, m 
+O\left(\frac{1}{r}\right) ,
\end{split}
\end{equation}
\begin{equation}
\begin{split}
\tilde n &= 
 \frac{\partial}{\partial \tilde u}  + O\left( \frac{1}{r}\right)
 =  \frac{1}{\dot\alpha}  \,  \frac{\partial}{\partial u}
+ O\left( \frac{1}{r}\right)\\
&=  \frac{1}{\dot\alpha}  \, n
+ O\left( \frac{1}{r}\right) ,
\end{split}
\end{equation}
\begin{equation}
\begin{split}
\tilde m &= 
\frac{\sqrt{2}\tilde P}{\tilde r}  \frac{\partial}{\partial \tilde \zeta}  
  + O\left( \frac{1}{r^2}\right) \\
 &= \frac{\sqrt{2}P_0 \tilde V\, w }{ r}  
\left( -\frac{\alpha_\zeta}{\dot\alpha} 
\,  \frac{\partial}{\partial u}
+ \frac{\partial}{\partial \zeta} \right)
+ O\left( \frac{1}{r^2}\right)\\
&=  -\frac{\sqrt{2}P_0 \tilde V\, w }{ r}  
\frac{\alpha_\zeta}{\dot\alpha} \,  n
+ \frac{\tilde V \, w}{V}m
+ O\left( \frac{1}{r^2}\right) ;
\end{split}
\end{equation}
since the metric expressed in terms of the new null tetrad 
must coincide with the metric
expressed in terms of the original null tetrad, it is deduced that
\begin{equation}\label{eq:tildeV}
\tilde V = \frac{V}{w}=\frac{V g_{ur}^0 }{\dot\alpha} ;
\end{equation}
therefore
\begin{equation}
\tilde m = m
  -\frac{\eth_V \alpha}{r \, \dot\alpha } \,  n
+ O\left( \frac{1}{r^2}\right) .
\end{equation}

The null tetrad transformation equations can be used to write the leading order
transformation relations for the spinor dyad associated to the 
null tetrad\cite{Geroch73,moreschi86}; namely
\begin{equation}
\tilde o^A = \sqrt{\dot\alpha} 
\left( o^A - \frac{\eth_{V} \alpha}{r\, \dot\alpha}\, \iota^A \right)
\end{equation}
and
\begin{equation}
\tilde \iota^A = \frac{1}{\sqrt{\dot\alpha}}\, \iota^A ;
\end{equation}
where $\eth_V$ is the edth operator of the metric (\ref{eq:desphere}).
Taking into account higher order transformations would include an equation of 
the form
\begin{equation}
  \label{eq:iotagenral}
  \tilde \iota^A = \frac{1}{\sqrt{\dot\alpha}}
  \left( \iota^A + h\, o^A \right);
\end{equation}
where in principle $h$ could be of order $O\left(r^0\right)$.

The regular dyad at future null infinity 
in terms of the spacetime one, can be given by
\begin{equation}
  \label{eq:omicronhat}
  \hat o^A = \Omega^{-1}\, o^A ,
\end{equation}
\begin{equation}
  \label{eq:iotahat}
  \hat \iota^A =  \iota^A .
\end{equation}

Then, the transformed regular dyad at future null infinity is given by
\begin{equation}
\begin{split}
\hat{\tilde o}^A &= \tilde\Omega^{-1}\, \tilde o^A = 
\frac{r}{w}\sqrt{\dot\alpha} 
\left( o^A - \frac{\eth_{V} \alpha}{r\, \dot\alpha}\, \iota^A \right) \\
&=\frac{1}{\sqrt{\dot\alpha}}
\left(\hat o^A - \frac{\eth_{V} \alpha}{\dot\alpha}\, 
\hat\iota^A \right)
,
\end{split}
\end{equation}

\begin{equation}
\hat{\tilde \iota}^A = \tilde \iota^A = \frac{1}{\sqrt{\dot\alpha}}\, 
\hat \iota^A ;
\end{equation}
where we are using $\Omega = \frac{1}{r}$.

\subsubsection{Transformation of $\Psi_4^0$}
We can now easily calculate the component $\Psi_4$ of the Weyl tensor, 
in leading orders,
with respect to the new null tetrad, obtaining
\begin{eqnarray}\label{eq:tpsi4}
\tilde \Psi_4^0 &=& \tilde{\Omega}^{-1} \Psi_{ABCD}
\hat{\tilde \iota}^A \hat{\tilde \iota}^B \hat{\tilde \iota}^C \hat{\tilde \iota}^D
= \frac{1}{w \dot\alpha^2} 
\Psi_4^0  .\\
&=& \frac{g_{ur}^0}{\dot\alpha^3} \Psi_4^0 
\end{eqnarray}

\subsubsection{Transformations to a Bondi system}\label{sec:trbondi}
Having analyzed how to transform among different frames we can now
consider our main task, to relate $\Psi_4$ calculated in an arbitrary
frame to that which would be obtained in a Bondi frame.
This frame satisfies, $g^{\tilde u \tilde r}=1$ and $\tilde V=1$ which implies
\begin{equation}
  \label{eq:dotalfa}
  \dot\alpha=V g_{ur}^0 \; ; \; \dot \alpha = w g_{ur}^0
\end{equation}
Which fix both $w$ and $\dot \alpha$ and indicates how
different quantities are to transform. In particular,
\begin{equation}\label{eq:tpsi4bondi}
\tilde \Psi_4^0 
= \frac{1}{V^3 (g_{ur}^0)^2} 
\Psi_4^0  ,
\end{equation}
Thus, knowledge of both $g_{ur}^0$ and $V$ allows one to obtain
$\tilde \Psi_4$ in terms of the more directly calculated $\Psi_4$.
We defer to section VI the discussion of how to obtain  $g_{ur}^0$ and $V$.

\subsection{Coordinate and tetrad transformations}\label{sec:generalasymptotic}
After examining the simpler case, we concentrate now on the general
case, namely we will include the possibility of $X^A_0 \neq 0$.
As before, the task at hand is to transform to an asymptotic Bondi
coordinate $(\tilde u, \tilde r, \tilde\zeta,\bar{\tilde\zeta})$
and tetrad frame $(\tilde \ell,\tilde n, \tilde m, \bar{\tilde m})$.
This transformation is of the form
\begin{align}
\tilde u &= \alpha(u,\zeta,\bar\zeta) + 
 \frac{\tilde u_1(u,\zeta,\bar\zeta)}{r}  \label{eq:tildeu2}
+ O\left(\frac{1}{r^2}\right) ,\\
\tilde r &= \frac{r}{w(u,\zeta,\bar\zeta)}  \label{eq:tilder2}
+ O\left(r^0\right)  ,\\
\tilde \zeta & = \tilde \zeta_0(u,\zeta,\bar\zeta) + O\left(\frac{1}{r}\right) .\label{eq:tildezeta2}
\end{align}
with $\dot \alpha >0$.

If one assumes that $\zeta$ is an stereographic coordinate
of the conformal unit sphere $(t=\text{const.}, r=\text{const.})$
it is only necessary to consider an angular transformation
of the form 
\begin{equation}
\tilde \zeta = \tilde \zeta_0(u,\zeta) + O\left(\frac{1}{r}\right) .
\label{eq:tildezeta2b}
\end{equation}
The stronger statement is as follows.

The stereographic coordinates of the sphere can be thought of as
a map of the extended complex plane.
In the asymptotic sphere defined by this setting, the 
angle transformation, at $u$=const., must be conformal.
But, a map of the extended complex plane
onto itself is conformal if and only if it is a Moebius transformation;
that is, a transformation of the form
\begin{equation}
\tilde \zeta = \frac{a \zeta + b}{c \zeta +d};
\end{equation}
with $ab - bc \neq 0$.

We proceed in two stages.\\

First, note that the contravariant metric components for the standard Bondi like
coordinate system is given by equations (3.13-18) of \cite{Moreschi87};
whose inverse is given by equations (3.19-24) of the same reference.
The only difference is that for the general tetrad here considered
\begin{equation}
g^{ur}=\frac{1}{g_{ur}} ;
\end{equation}
with $g_{ur}$ not necessarily being equal to $1$.
In order to deduce the required relations we start from the null
tetrad defined by (\ref{uno}),
\begin{equation}
\left(\ell^a\right) = 
\left(A \frac{\partial}{\partial r}
\right)^a ,
\end{equation}
(\ref{eq:vecm}) and (\ref{eq:vecn}). Then the contravariant components of the
metric (the inverse metric), is given by
\begin{align}
g^{uu} &= 0 ,\\
g^{ur} &= A ,\\
g^{ui} &= 0 ,\\
g^{rr} &= 2 U ,\\
g^{ri} &= X^i ,\\
g^{ij} &= -(\xi^i \bar\xi^j + \bar\xi^i \xi^j) .
\end{align}
Then the metric is given by
\begin{align}
g_{ur} &= \frac{1}{A} ,\\
g_{rr} &= 0 ,\\
g_{ri} &= 0 ,\\
g_{uu} &= -2 \frac{U}{A} + X^i X^j g_{ij}  ,\\
g_{ui} &= - g_{ij} X^j ,\\
g_{ij} &= (g^{ij})^{-1} = 
  -d \epsilon_{ik} \epsilon_{jl}(\xi^i \bar\xi^j+ \bar\xi^i \xi^j) ;
\end{align}
with $i,j,k,l=2,3$, $d = \det(g_{ij})$,
$\epsilon_{ij} = -\epsilon_{ji}$ and $\epsilon_{23}=1$.
In particular, defining the quantity
\begin{equation}
\lambda = \epsilon_{ij} \xi^i \bar\xi^j ;
\end{equation}
one has that
\begin{equation}
d = \frac{1}{|\lambda|^2} .
\end{equation}

We will be interested in calculating the transformation
of the radiation field component $\Psi_4^0$;
which depends on the transformation properties of
the dyad spinor $\iota^A$. This can be read-off
from the transformation properties of the tetrad vector $n$.
A straightforward calculation reveals that the one form $n_a$ is
\begin{equation}
\begin{split}
(n_a) &= (g_{ab} n^b) =
  g_{ur} dr + g_{uu} du + g_{ui} dx^i \\
& \quad\quad + U g_{ru} du
+ X^i g_{iu} du + X^i g_{ij} dx^j \\
&= \frac{1}{A} dr \\
&\;+ \left(
-2 \frac{U}{A} + X^i X^j g_{ij}
+ U \frac{1}{A} + X^i (- g_{ij} X^j)
\right) du \\
&\; +\left(
- g_{ij} X^j + X^j g_{ij}
\right) dx^i \\
&= \frac{1}{A} \left( dr - U du \right) .
\end{split}
\end{equation}

In the Bondi frame the following relations are satisfied,
\begin{align}
(\tilde \ell_a) &=  d\tilde u , \\
(\tilde m_a) &=  \frac{1}{\tilde \lambda} \epsilon_{ij} \tilde\xi^i \tilde X^j d\tilde u
- \frac{1}{\tilde \lambda} \epsilon_{ij} \tilde \xi^i d\tilde x^j ,  \\
(\tilde n_a) &=  d\tilde r - \tilde U d\tilde u .
\end{align}

Additionally, in terms of the original coordinate system, the one-forms
associated to the Bondi coordinates, to leading order, are given by
\begin{equation}
d\tilde u = \dot\alpha \, du
+ \alpha_\zeta  \, d\zeta + \alpha_{\bar\zeta}  \, d\bar\zeta 
+O\left(\frac{1}{r}\right) ,
\end{equation}
\begin{equation}
d\tilde r = \frac{1}{w} \, dr - \frac{r}{w^2}
\left(
\dot w du + w_\zeta d\zeta +w_{\bar\zeta} d\bar\zeta
\right)
+O(r^0) ,
\end{equation}
and
\begin{equation}
d\tilde \zeta = \dot{\tilde \zeta} \, du
+ \tilde \zeta_\zeta  \, d\zeta + \tilde \zeta_{\bar\zeta}  \, d\bar\zeta 
+O\left(\frac{1}{r}\right) .
\end{equation}

Therefore from the contribution to the metric of
$\tilde\ell_a \tilde n_b$, one deduces that
\begin{equation}
\dot\alpha \frac{1}{w} = g_{ur}^0 
;
\end{equation}
that is $w$ is determined by the choice of $\alpha$ and the 
value of $g_{ur}^0$.

In order to study the asymptotic transformation of the spinors
$\tilde\iota^A$, we examine first the tetrad one-form $\tilde n_a$.

\begin{equation}
\begin{split}
 d\tilde r - \tilde U d\tilde u = &
 \frac{1}{w} \, dr - \frac{r}{w^2}
\left(
\dot w du + w_\zeta d\zeta +w_{\bar\zeta} d\bar\zeta
\right)
+O(r^0) \\
&-\tilde U
\left(
\dot\alpha \, du
+ \alpha_\zeta  \, d\zeta + \alpha_{\bar\zeta}  \, d\bar\zeta 
+O\left(\frac{1}{r}\right)
\right) \\
=& \frac{1}{w} 
\left(
dr
-
\left(
\frac{r}{w} \dot w +\tilde U \dot\alpha
\right) du
\right) \\
& +\frac{r}{w^2}
\left(
 w_\zeta d\zeta +w_{\bar\zeta} d\bar\zeta
\right) + (\text{lower orders}) .
\end{split}
\end{equation}

It is important to note that in the general case,
the vector $n^a$, in the asymptotic region, 
will have a non zero component in the direction of the
spacelike direction given by the conformal Bondi vectors
$\hat{\tilde m}$ and $\hat{\bar{\tilde m}}$.
This can be seen from the fact that regular tetrad (with hat)
at future null infinity is related to the standard tetrad
in the asymptotic region by
\begin{align}
\hat \ell_a &= \ell_a , \\
\hat m_a &= \Omega \,m_a , \\
\hat n_a &= \Omega^2 \, n_a ,
\end{align}
and
\begin{align}
\hat \ell^a &= \Omega^{-2} \, \ell^a , \\
\hat m^a &= \Omega^{-1} \, m^a , \\
\hat n^a &=  n^a ;
\end{align}
where $\Omega$ is the conformal factor that defines
the regular metric $\hat g_{ab} = \Omega^2 \,{g}_{ab}$.
Then, since in the asymptotic region
\begin{equation}
({\tilde m}_a) = \frac{\tilde r}{\sqrt{2} \tilde P_0 \tilde V} 
d\bar{\tilde \zeta} + O(\tilde r^0) ,
\end{equation}
one can deduce
\begin{equation}
\hat n^a (\hat{\tilde m}_a) = \frac{1}{\sqrt{2} \tilde P_0 \tilde V}
\left(
\dot{\bar{\tilde \zeta}} + X^{\bar \zeta} \;\bar{\tilde \zeta}_{\bar \zeta}
\right)
 + O(\tilde r^{-1}) ;
\end{equation}
where $\Omega = \frac{1}{\tilde r}$ in this case.
It is important here to realize the following.
Extending the hypersurface $u=$const. in the asymptotic
region, one defines a section $S$ of future null infinity;
in which by construction the vectors 
$\Omega^{-1}m^a$ and $\Omega^{-1}\bar m^a$ 
are tangent. Also, the regular extension of $\ell$, namely
$\Omega^{-2} \ell^a$, is orthogonal to $S$. But, since
future null infinity is a null hypersurface, one has that
$n^a$, which is a null vector orthogonal to $S$,
must be tangent to future null infinity.
Since that regular extension of the Bondi frame to
future null infinity is such that $\Omega^{-1} \tilde m^a$
and $\Omega^{-1} \bar{\tilde m}^a$
will be tangent to $S$; one deduces that the function
$\zeta_0$ must be chosen such that
\begin{equation}
\dot{\bar{\tilde \zeta}} + X^{\bar \zeta} \;\bar{\tilde \zeta}_{\bar \zeta}
= 0 ; \label{eq:condzeta0}
\end{equation}
since the vector $\hat{\tilde m}_a$ must be orthogonal
to $n^a$ at future null infinity.\\

Next, let us study the asymptotic form of $\tilde n^a$.
Let us first note that
\begin{equation}
\frac{\partial}{\partial \tilde u} =
\frac{\partial u}{\partial \tilde u}\frac{\partial}{\partial u}
+\frac{\partial \zeta}{\partial \tilde u}\frac{\partial}{\partial \zeta} 
+\frac{\partial \bar\zeta}{\partial \tilde u}\frac{\partial}{\partial \bar\zeta}
;\label{eq:dutildeu}
\end{equation}
where the last two terms are different from zero, since one has
\begin{equation}
0=\frac{\partial \tilde\zeta}{\partial \tilde u} =
\frac{\partial u}{\partial \tilde u}\frac{\partial \tilde \zeta}{\partial u}
+\frac{\partial \zeta}{\partial \tilde u}\frac{\partial \tilde \zeta}{\partial \zeta} .\label{eq:dtildezdtildeu}
\end{equation}
Applying (\ref{eq:dutildeu}) onto $\tilde u$ and using
(\ref{eq:dtildezdtildeu}) one can see that
\begin{equation}
\frac{\partial u}{\partial \tilde u}
= \frac{1}{\dot \alpha 
- \alpha_\zeta \frac{\dot{\tilde {\zeta}}_0}{\tilde {\zeta_0}_\zeta}
- \alpha_{\bar\zeta} \frac{\dot{\bar{\tilde \zeta}}_0}{{\bar{ \tilde \zeta}_0}_{\bar\zeta}}} 
,
\end{equation}
and
\begin{equation}
\frac{\partial \zeta}{\partial \tilde u} = 
- 
\frac{1}{\dot \alpha 
- \alpha_\zeta \frac{\dot{\tilde {\zeta}}_0}{\tilde {\zeta_0}_\zeta}
- \alpha_{\bar\zeta} \frac{\dot{\bar{\tilde \zeta}}_0}{{\bar{ \tilde \zeta}_0}_{\bar\zeta}}}
\frac{\dot{\tilde {\zeta}}_0}{\tilde {\zeta_0}_\zeta}
+ O\left( \frac{1}{r}\right) . 
\end{equation}
Similarly. one has
\begin{equation}
\frac{\partial \bar\zeta}{\partial \tilde u} = 
- 
\frac{1}{\dot \alpha 
- \alpha_\zeta \frac{\dot{\tilde {\zeta}}_0}{\tilde {\zeta_0}_\zeta}
- \alpha_{\bar\zeta} \frac{\dot{\bar{\tilde \zeta}}_0}{{\bar{ \tilde \zeta}_0}_{\bar\zeta}}}
\frac{\dot{\bar{\tilde \zeta}}_0}{{\bar{\tilde \zeta}_0}_{\bar\zeta}} 
+ O\left( \frac{1}{r}\right) \; ;
\end{equation}
and thus,
\begin{equation}
\frac{\partial}{\partial \tilde u} =
\frac{1}{\dot \alpha 
- \alpha_\zeta \frac{\dot{\tilde {\zeta}}_0}{\tilde {\zeta_0}_\zeta}
- \alpha_{\bar\zeta} \frac{\dot{\bar{\tilde \zeta}}_0}{{\bar{ \tilde \zeta}_0}_{\bar\zeta}}}
\left(
\frac{\partial}{\partial u}
-\frac{\dot{\tilde \zeta}}{\tilde \zeta_\zeta}
\frac{\partial}{\partial \zeta}
-\frac{\dot{\bar{\tilde \zeta}}}{\bar{\tilde \zeta}_\zeta}
\frac{\partial}{\partial \bar\zeta}
\right) .
\end{equation}

To leading orders in the asymptotic tetrad transformation, $\tilde n$ and $n$
are related by
\begin{equation}
\begin{split}
\tilde n &= 
 \frac{\partial}{\partial \tilde u}  + O\left( \frac{1}{r}\right)
\\
&=  
\frac{1}{\dot \alpha 
- \alpha_\zeta \frac{\dot{\tilde {\zeta}}_0}{\tilde {\zeta_0}_\zeta}
- \alpha_{\bar\zeta} \frac{\dot{\bar{\tilde \zeta}}_0}{{\bar{ \tilde \zeta}_0}_{\bar\zeta}}}
\left(
\frac{\partial}{\partial u}
-\frac{\dot{\tilde \zeta}}{\tilde \zeta_\zeta}
\frac{\partial}{\partial \zeta}
-\frac{\dot{\bar{\tilde \zeta}}}{\bar{\tilde \zeta}_\zeta}
\frac{\partial}{\partial \bar\zeta}
\right)
+ O\left( \frac{1}{r}\right) \\
&= 
\frac{1}{\dot \alpha 
- \alpha_\zeta \frac{\dot{\tilde {\zeta}}_0}{\tilde {\zeta_0}_\zeta}
- \alpha_{\bar\zeta} \frac{\dot{\bar{\tilde \zeta}}_0}{{\bar{ \tilde \zeta}_0}_{\bar\zeta}}}
\left(
n 
- X_0^\zeta \frac{\partial}{\partial \zeta}
- X_0^{\bar\zeta} \frac{\partial}{\partial \bar\zeta}
\right. \\
&\left.\qquad\qquad\qquad\qquad
-\frac{\dot{\tilde \zeta}}{\tilde \zeta_\zeta}
\frac{\partial}{\partial \zeta}
-\frac{\dot{\bar{\tilde \zeta}}}{\bar{\tilde \zeta}_\zeta}
\frac{\partial}{\partial \bar\zeta}
\right)
+ O\left( \frac{1}{r}\right) \\
&=
\frac{1}{\dot \alpha 
- \alpha_\zeta \frac{\dot{\tilde {\zeta}}_0}{\tilde {\zeta_0}_\zeta}
- \alpha_{\bar\zeta} \frac{\dot{\bar{\tilde \zeta}}_0}{{\bar{ \tilde \zeta}_0}_{\bar\zeta}}}
\;
n + O\left( \frac{1}{r}\right)
,
\end{split}
\end{equation}
since by virtue of (\ref{eq:condzeta0}) and its complex conjugate,
the asymptotic
terms involving spatial directions cancel.

Additionally,
\begin{equation}
0=\frac{\partial \tilde u}{\partial \tilde \zeta} =
\frac{\partial u}{\partial \tilde \zeta}\frac{\partial \tilde u}{\partial u}
+\frac{\partial \zeta}{\partial \tilde \zeta}\frac{\partial \tilde u}{\partial \zeta} 
=\dot \alpha \frac{\partial u}{\partial \tilde \zeta}
+ \alpha_\zeta \frac{\partial \zeta}{\partial \tilde \zeta}
;
\end{equation}
thus
\begin{equation}
\frac{\partial \zeta}{\partial \tilde \zeta}
=
-\frac{\dot \alpha}{\alpha_\zeta}
\frac{\partial u}{\partial \tilde \zeta} .
\end{equation}

Last, from
\begin{equation}
1=\frac{\partial \tilde \zeta}{\partial \tilde \zeta} =
\frac{\partial u}{\partial \tilde \zeta}
\frac{\partial \tilde \zeta}{\partial u}
+\frac{\partial \zeta}{\partial \tilde \zeta}
\frac{\partial \tilde \zeta}{\partial \zeta} 
=\dot{\tilde \zeta} \frac{\partial u}{\partial \tilde \zeta}
+ \tilde\zeta_\zeta \frac{\partial \zeta}{\partial \tilde \zeta}
=
\dot{\tilde \zeta} \frac{\partial u}{\partial \tilde \zeta}
- \tilde\zeta_\zeta 
\frac{\dot \alpha}{\alpha_\zeta}
\frac{\partial u}{\partial \tilde \zeta}
;
\end{equation}
one has
\begin{equation}
\frac{\partial u}{\partial \tilde \zeta}
=
\frac{\alpha_\zeta}{\alpha_\zeta \; \dot{\tilde \zeta} 
-\dot \alpha \;\tilde\zeta_\zeta} ;
\end{equation}
and
\begin{equation}
\frac{\partial \zeta}{\partial \tilde \zeta}
=
-\frac{\dot \alpha}
{\alpha_\zeta \; \dot{\tilde \zeta} 
-\dot \alpha \;\tilde\zeta_\zeta}
.
\end{equation}

The asymptotic behavior of the vector $\tilde m$ is therefore
\begin{equation}
\begin{split}
\tilde m &= 
\frac{\sqrt{2}\tilde P}{\tilde r}  \frac{\partial}{\partial \tilde \zeta}
  + O\left( \frac{1}{r^2}\right) \\
 &= \frac{\sqrt{2}P_0 \tilde V\, w }{ r}  
\left( 
-\frac{\alpha_\zeta}
{\dot\alpha\;\tilde\zeta_\zeta-\alpha_\zeta \; \dot{\tilde \zeta}}
\,  \frac{\partial}{\partial u}
+\frac{\dot \alpha}
{\dot \alpha \;\tilde\zeta_\zeta
-
\alpha_\zeta \; \dot{\tilde \zeta}}
\frac{\partial}{\partial \zeta} 
\right) \\
&\qquad\qquad+ O\left( \frac{1}{r^2}\right)\\
&=  
-\frac{\sqrt{2}P_0 \tilde V\, w }{ r} \;
\frac{\alpha_\zeta}{\dot\alpha\;\tilde\zeta_\zeta
-
\alpha_\zeta \; \dot{\tilde \zeta}} \,  n
+ \frac{\tilde V \, w}{V}
\;
\frac{\dot \alpha}
{\dot \alpha \;\tilde\zeta_\zeta
-
\alpha_\zeta \; \dot{\tilde \zeta}}
m \\
&\qquad\qquad
+ O\left( \frac{1}{r^2}\right) .
\end{split}
\end{equation}
Since the metric expressed in terms of the new null tetrad 
must coincide with the metric
expressed in terms of the original null tetrad, it is deduced that
\begin{equation}\label{eq:tildeVb}
1=\tilde V 
=
\frac{V |\dot \alpha \;\tilde\zeta_\zeta
-
\alpha_\zeta \; \dot{\tilde \zeta}|}
{w \, \dot \alpha}
=
\frac{V g_{ur}^0|\dot \alpha \;\tilde\zeta_\zeta
-
\alpha_\zeta \; \dot{\tilde \zeta}|}
{\dot \alpha^2}
;
\end{equation}
where we have used that the tilde system  is Bondi like.


One then can deduce the transformation
\begin{equation}
\tilde \iota^A = \frac{1}
{\sqrt
{\dot \alpha 
- \alpha_\zeta \frac{\dot{\tilde {\zeta}}_0}{\tilde {\zeta_0}_\zeta}
- \alpha_{\bar\zeta} \frac{\dot{\bar{\tilde \zeta}}_0}{{\bar{ \tilde \zeta}_0}_{\bar\zeta}}}
}
\, \iota^A .
\end{equation}

Consequently the radiation component $\Psi_4^0$ transforms in the 
following way
\begin{equation}\label{eq:tpsi4b}
\begin{split}
\tilde \Psi_4^0 &= \tilde{\Omega}^{-1} \Psi_{ABCD}
\hat{\tilde \iota}^A \hat{\tilde \iota}^B \hat{\tilde \iota}^C \hat{\tilde \iota}^D \\
&= \frac{1}{w}
\frac{1}
{\left(
{\dot \alpha 
- \alpha_\zeta \frac{\dot{\tilde {\zeta}}_0}{\tilde {\zeta_0}_\zeta}
- \alpha_{\bar\zeta} \frac{\dot{\bar{\tilde \zeta}}_0}{{\bar{ \tilde \zeta}_0}_{\bar\zeta}}}
\right)^2
}
\Psi_4^0 \\
&=
\frac{g_{ur}^0}
{\dot \alpha \left(
{\dot \alpha 
- \alpha_\zeta \frac{\dot{\tilde {\zeta}}_0}{\tilde {\zeta_0}_\zeta}
- \alpha_{\bar\zeta} \frac{\dot{\bar{\tilde \zeta}}_0}{{\bar{ \tilde \zeta}_0}_{\bar\zeta}}}
\right)^2
}
\Psi_4^0
.
\end{split}
\end{equation}
Equivalently, this can be expressed as
\begin{equation}
\tilde \Psi_4^0 =
\frac{g_{ur}^0}
{\dot \alpha \left(
{\dot \alpha 
+ \alpha_\zeta X_0^{\zeta}
+ \alpha_{\bar\zeta} X_0^{\bar \zeta}}
\right)^2
}
\Psi_4^0 \label{eq:psi4c}
\end{equation}

Let us emphasize that $\tilde\zeta_0$ is chosen to make $\tilde X_0 = 0$;
while $\alpha$ is chosen to make $\tilde V = 1$.


This complicated transformation of the radiation field indicates the
convenience of adapting the numerical code such that
$X^i = 0 +O(r^{-1})$; so that $\dot{\tilde\zeta} = 0$. Otherwise
the determination of $\alpha$, needed in (\ref{eq:tpsi4b})
or (\ref{eq:psi4c}), would be very difficult.

\section{Examples}
\subsection{Conventions and formulae. Factors of $2$}
In section I we presented the standard expressions for the radiated
four-momentum. Yet, there are differences in factors with
the definitions employed in numerical implementations. The difference
arises from the frame adopted for the extraction procedure. The analysis
presented in this work can be used to clarify these differences.
In a Bondi frame, the expression for the radiated four-momentum $P^a$ is,

\begin{equation}
\dot P^a =  - \frac{1}{4 \pi } \int_S \hat l^{a} \dot \sigma^0 \dot {\bar \sigma}^0 dS^2 .
\end{equation}
(with $\dot{}$ denoting a derivative with respect to $\tilde u$)
This, can be re-expressed in terms of $\Psi_4^0$ by employing the identity
valid in a Bondi frame $\tilde \Psi_4^0 = - \ddot {\bar \sigma}^0$. Where we have
used the $\tilde{} $ to denote a quantity obtained in a Bondi frame (we don't
do that for $\sigma^0$ as here we only use it as obtained in the Bondi frame).
 
\begin{equation}
\dot P^a =  - \frac{1}{4 \pi } \int_S {\hat l^{a}} 
\left| \int_{-\infty}^{\tilde u} \tilde \Psi_4^0 d\tilde u' \right|^2 dS^2 .
\end{equation}

In a numerical implementation, as discussed, the readily available
$\Psi_4^0$ differs from the Bondi one by
\begin{equation}
\tilde \Psi_4 = \frac{\Psi_4}{(g_{ur}^0)^2 V^3} .
\end{equation}
Thus
\begin{equation}
\dot P^a =  - \frac{1}{4 \pi } \int_S {\hat l^{a}} 
\left| \int_{-\infty}^{\tilde u} \frac{\Psi_4}{(g_{ur}^0)^2 V^3} d\tilde u' \right|^2 dS^2 .
\end{equation}

\subsubsection{Standard approach, boost ambiguity and resulting factors of $2$}
We recall that the null tetrads are not uniquely defined as they have the boost
freedom $\ell^a \rightarrow \ell^a \lambda^{-1}, n^a \rightarrow n^a \lambda$ and spin freedom 
$m^a \rightarrow m^a e^{i\gamma}$\cite{Geroch73}. This freedom can bring about additional factors in the resulting
expression for the Weyl scalars, fortunately a Bondi frame naturally fixes this ambiguity. In what
follows we illustrate how this freedom is fixed by ensuring a Bondi tetrad frame is adopted in a simple
exmaple. This, in passing, will make evident the different factors encountered in commonly employed formulae. 
In the extraction procedure it has become custommary to introduce a tetrad frame in the following form \cite{lazarus}. 
First an extraction worldtube
is defined as the Cartesian $x^2+y^2+z^2=R^2$.
Then, three vectors (labeled by $J$) $\tilde v_J^i$ at a hypersurface slice are adopted in the following way:
\begin{equation*}
\tilde v_1^i = (-y,x,0), \tilde v_2^i = (x,y,z)^i, \tilde v_3^i = det(\gamma)^{(1/2)} \gamma^{ij} \epsilon_{jlm} v_1^l v_2^m;
\end{equation*}
where $\gamma_{ij}$ is the induced metric on the spacelike hypersurface at a given time.
In our present case, this reduces to 
\begin{equation}
\tilde v_1^i = \partial_{\phi}^i, \tilde v_2^ i =\partial_{r}^i, \tilde v_3^i = \partial_{\theta}^i.
\end{equation}
The next step involves a Gram-Schmidt procedure to construct three orthonormal vectors $v_J^i$ with respect
to $\gamma_{ij}$. Finally four spacetime vectors are easily constructed by
$r^a=(0,v_2); \theta^a = (0,v_3) ; \phi^a = (0,v_1)$, which together with $N^a$ (the unit timelike vector normal to the
hypersurface) can be employed to construct  the tetrad as,
\begin{eqnarray*}
\ell'^a &=& \frac{1}{\sqrt{2}} (N^a + r^a)  \\
n'^a  &=&  \frac{1}{\sqrt{2}} (N^a - r^a)  \\
m'^a &=& \frac{1}{\sqrt{2}} (\theta^a + i \phi^a) ,
\end{eqnarray*}
with $r^2 = x^2+y^2+z^2$.
Consider the simplest case of a spacetime whose metric, 
is given by $g_{ab} = \eta_{ab} + h_{ab}$ with $h_{ab}$ a sufficiently
fast decaying functions at far distances and $\eta_{ab}$ a flat metric
in Cartesian coordinates.
In this case, to leading order, the tetrad resulting from the standard 
procedure is,
\begin{eqnarray}
\ell'^a &=& \frac{1}{\sqrt{2}} (\partial_t^a + \partial_R^a) \, , \\
n'^a &=& \frac{1}{\sqrt{2}} (\partial_t^a -  \partial_R^a) \, , \\
m'^a &=& \frac{1}{r \sqrt{2}} \left (\partial_{\theta}^a +  
\frac{i}{\sin(\theta)} \partial_{\phi}^a \right) \, .
\end{eqnarray}
The main difference with the previous tetrads is that in the former
one has used a null coordinate $u$ that in this case would be
of the form $u= t - R$; with the first null tetrad vector
given by $\ell_a = du_a$. Therefore, one has 
$\ell'^a = \frac{1}{\sqrt{2}} \ell^a$ in this case and so
$\iota'^A = 2^{\frac{1}{4}} \iota^A$. Then, in the calculation
of the radiation component of the Weyl tensor one would have
$\Psi'_4 = 2 \Psi_4$. 
The expression for the radiated momentum in these coordinates results, 
\begin{equation}
\dot P^a =  - \frac{1}{16 \pi } \int_S  \hat l^{a} 
\left| \int_{-\infty}^{u} {\Psi_4'} d\tilde u \right|^2 dS^2 .
\end{equation}
This then explains the factor of $4$ difference
in the expressions cited in \cite{bruegman,loustolazzarus,nasawaves} 
with that in eqn (\ref{radiate}) (and those in the standard literature \cite{newmantod,sources}, etc.).

\subsection{Teukolsky waves and wave extraction}
As a second example, we adopt linearized waves on flat spacetime as given
by the so called Teukolsky waves\cite{teukolskywaves}. We adopt this example,
coupled with a possible coordinate transformation to illustrate the differences
that may arise when suitable contact with a Bondi frame is missing.
We adopt the simplest expression for the line element given
by
\begin{eqnarray}
ds^2 &=& dt^2 - (1+A f_{rr}) dr^2 - 2 B f_{r\theta} dr d\theta \nonumber \\
& & - (1 + 3 C f^1_{\theta\theta} - A ) r^2 d\theta^2 \nonumber \\
& & -  \left (1 + 3 C f^1_{\phi\phi}  
 - A f^2_{\phi\phi} \right ) r^2 \sin(\theta)^2 d\phi^2
\end{eqnarray}
where
\begin{eqnarray*}
f_{rr} = 2 - 3 \sin(\theta)^2\; \; ; \; \; f_{r\theta} &=& -3 \sin(\theta) \cos(\theta) \\
f^1_{\theta\theta} = -f^1_{\phi\phi} = f^2_{\phi\phi} +  1 &=& 3 \sin(\theta)^2.
\end{eqnarray*}
and
\begin{eqnarray}
A &=& 3 \left ( \frac{F^{(2)}}{r^3} + \frac{3F^{(1)}}{r^4} + \frac{3F}{r^5} \right) \\
B &=& - \left ( \frac{F^{(3)}}{r^2} + \frac{3F^{(2)}}{r^3} + \frac{6F^{(1)}}{r^4} + \frac{6F}{r^5} \right) \\
C &=& \frac{1}{4} \left (\frac{F^{(4)}}{r} + \frac{2F^{(3)}}{r^2} + \frac{9F^{(2)}}{r^3} \right . \nonumber \\
  &~& {} \; \; \; \; \; +  \left. \frac{21F^{(1)}}{r^4} + \frac{21F}{r^5} \right)
\end{eqnarray}
with $F=F(t-r)$, $F^{(n)} =\left( \frac {d^n F(x)}{dx^n} \right)_{x=t-r}$.
 
The Riemann tensor for such a line element can be straightforwardly constructed. Since 
to leading order the line element is just the flat metric, it is straightforward
to identify the Bondi frame. A simple calculation gives,
\begin{equation}
\tilde \Psi_4^0 = \frac{3}{8} \sin(\theta)^2 F^{(6)}(t-r) \label{teukbondipsi4}
\end{equation} 
On the other hand, we can calculate $\Psi_4$ with the standard procedure. 
We consider two cases: (I) calculating $\Psi_4$ in the coordinates $(t,r,\theta,\phi)$
and (II) considering the transformation $\tilde r \rightarrow r g(t)$ which
induces a non-trivial $V(t)$. The induced tetrad, to the order that
enters in the calculation of the leading order of $\Psi_4$ are
\begin{eqnarray*}
\ell^a &=& \frac{1}{\sqrt{2}} \left ( \partial_t^a + \frac{1}{g} (1 - r \dot g) \partial_R^a\right ) \, ,  \\
n^a  &=&  \frac{1}{\sqrt{2}} \left ( \partial_t^a - \frac{1}{g} (1 - r \dot g) \partial_R^a\right ) \, , \\
m^a &=& \frac{1}{r g \sqrt{2}} (\partial_{\theta}^a + i \partial_{\phi}^a) .
\end{eqnarray*}
Taking $g \equiv 1$ gives the tetrad for case (I).
A straightforward calculation gives this case,
\begin{eqnarray}
\Psi_4^0 = \frac{3}{4} {F^{(6)}( t - r)} \sin(\theta)^2 \, ;
\end{eqnarray}
while for case (II)
\begin{eqnarray}
\Psi_4^0 = \frac{3}{4} \frac{F^{(6)}(t - r g) \sin(\theta)^2}{g} \, .
\end{eqnarray}

Notice a difference of a factor $g^{-1}$ appearing in case (II) in addition to a factor of $2$ when
compared to eqn (\ref{teukbondipsi4}). These factors  result from $V=g^{-1}$ and $g_{ur}^0=\sqrt{2} g$.
The corrections described in section IV can be used to reconcile these differences and obtain
Bondi's expression in a way mostly independent of gauge issues.
In the next
section we discuss how the different factors involved can be obtained in generic settings.

\section{Obtaining the missing links}

\noindent
From our previous discussion, it is clear that one must take into account the effects
caused by the conformal factor $V$ and the induced metric
components $g_{ur},g_{ui}$. These later factors can be read-off
from the expression of the null one-form $\tilde n_a$ in the induced null coordinates
at $\Gamma$. Namely $l_a = du_a$ defines the function $u$ and the transformation
from the $(t,r,\zeta,\bar \zeta)$ to the $(u,r,\zeta, \bar \zeta)$ coordinates can be employed
to express $n_a$ in this system. Then, $g_{ur}^{-1} = l^a dr_a$; $X^i = n^a dx^{i}_a$.

The extraction of the conformal factor $V$ involves more work but can be obtained 
by considering a conformal transformation
of the sphere and evaluating its curvature scalar\cite{hpgn}. 
Recall that any metric on $S^{2}$ is conformally related to that of the unit sphere, thus
it must be the case that $\hat g_{ij} \equiv g_{ij} R^{-2} = \omega^{-2} q_{ij}$. 
Thus, the unit sphere metric and the induced metric at the worldtube
at a constant time are related by a conformal factor $\omega$. With our
conventions then $V=\omega$.
A way to solve for this factor is to compute the scalar curvature
of the cut $S$ on the worldtube. This gives rise to the following
relation
\begin{equation}
{\cal R} = 2 \left ( V^2 + D^A D_A \ln V \right ),
\end{equation}
where ${\cal R}$ is the Ricci scalar curvature and $D_i$ is the
covariant derivative of the metric $\hat g_{ij}$.
Let us observe that in two dimensions one has ${\cal R} = 2 K_V$.

This equation can be solved to obtain $V$. Notice that this equation
admits more than one solution due to the rotational group of transformations.
One way to solve it, and implicitly fix this freedom is motivated in techniques 
introduced to find apparent horizons in numerical simulation\cite{Carsten,nakamura}. 
In this approach we express $V$ (or a related quantity) in terms of a spherical-harmonic expansion, and then
obtain a recursive relation to solve for the expansion coefficients.
We begin by considering\footnote{--Alternatively, one could have expressed $V$ 
itself in terms of a spectral series--.}
\begin{equation}
W \equiv V^2 = \Sigma_{lm} a_{lm} Y_{lm}
\end{equation}
with $Y_{lm}$ spherical harmonics satisfying $\bar\nabla^2 Y_{lm} = - l (l+1) Y_{lm}$ with
$\bar\nabla^2$ the Laplacian operator with respect to the unit sphere metric expressed
in the same coordinates as those where $\hat g_{ij}$ is known 
(we will refer to this as $\bar q_{ij}$).
The solution we seek satisfies $H(W) = 0$ with
\begin{equation}
H= 2 W + D^i D_i \ln W  - {\cal R}.
\end{equation}
Next, we consider the equation
\begin{equation}
\rho H(W) + \bar\nabla^2 W = \bar\nabla^2 W \label{conformalequation}
\end{equation}
with $\rho$ a function to be determined in a suitable manner. The solution we seek
turns the equation above into an identity.
Consider now integrating this equation over the sphere having replaced our anzats for $W$.
This will provide a recursion relation for obtaining the parameters $a_{lm}$ (for $l\neq 0$) as,
\begin{equation}
\int_S \bar Y_{lm} (\rho H + \bar\nabla^2 W) d\Omega = -l (l+1) a_{lm} \label{integral}
\end{equation}
Thus, starting with trial values for $\{a_{lm}\}$ a new set can be obtained through the equation
above. Two ingredients remain to be provided however, one is how $a_{00}$ is to be set and the other
a plausibility argument for the convergence of the method.
We describe first the latter issue and then discuss the former. This argument relies
on concentrating on the principal part of the system $(\rho H + \bar\nabla^2 W)$ and to exploit
$\rho$ to our advantage.
Inspection of these terms allows one to re-express them as $M^{ij} \bar D_i \bar D_j W + S$ with
\begin{eqnarray}
M^{ij} &=& \frac{\rho}{W} \hat g^{ij} + \bar q^{ij} \\
S &=& \rho \left ( 2 W - \frac{1}{W} \hat g^{ij} C^k_{ij} \bar D_k W \nonumber \right . \\
  &  & - \left . \frac{1}{W^2} \hat g^{ij}\bar D_i W \bar D_j W -{\cal R} \right ).
\end{eqnarray}
with $\bar D$ the covariant derivative associated with $\bar q_{ij}$ and $C^k_{ij}$ the tensor
relating the connections $\bar D_i$ with $D_i$.
Notice now that equation (\ref{conformalequation}) can be formally solved by the iteration scheme
\begin{equation}
W^{(n+1)} = (\bar\nabla^2)^{-1} (\rho H + \bar\nabla^2) W^{(n)}
\end{equation}
Here the standard argument applies, the Laplacian operator smoothes out high frequencies and by properly
adjusting $M^{ij}$ such that the highest order derivatives are removed the successive solutions
will be smoother which is a requirement for convergence. The factor $\rho$ can be chosen so as to
cancel particular terms in the second-order derivatives, or try to cancel as much as possible
$M^{ij}$ (see the discussion in \cite{Carsten}.
For instance, if $\bar q^{ij} \simeq q^{ij}$ the trivial choice $\rho = -1$ is ideally suited for
this purpose.

Last we comment on how $a_{00}$ is obtained. At the first step of the iteration, a useful
choice is induced by assuming $W^{(1)} = a_{00}^1 Y_{00}$. Then, one simply has
\begin{equation}
0 = \int (2 W^{(1)} - {\cal R} )  Y_{00}  d\Omega = 2 a_{00}^1 -  \int {\cal R} Y_{00} d\Omega
\end{equation}
At subsequent steps, having fixed $a_{lm}^{n+1}$ for $l>0$, equation
(\ref{integral}) can be employed to fix $a_{00}^{n+1}$. However, due to the non-linearities of the equation
this might turn out to be difficult.  A simpler, more direct route, is to do so from
\begin{equation}
\int_S \left (  ( 2 W^{(n+1)} ) +  D^i D_i \ln W^{(n)} - {\cal R} \right ) Y_{00} d\Omega = 0
\end{equation} 

Which can be used to determine $a_{00}^{n+1}$ after all other coefficients have been obtained.

\section{Final comments}
In this work we have re-examined the issue of computing
gravitational radiation effects through the use of Weyl scalars.
The analysis reveals which correcting factors are to be accounted
for if the coordinates adopted do not conform to a Bondi system.
To date while these corrections have not been taken into account in numerical
efforts the waveforms obtained both appear quite reasonable and, most
importantly, agreeing across different implementations.
What is then the expected contribution of the corrections 
indicated here or the relevance of the present discussion in light of
these observations? \\

First, at least the formalism presented
in this work allows for easily enforcing consistency among different implementations.
Namely, while coordinates typically used vary among different efforts,
considering the transformation to a Bondi system provides a common frame for
the computation. As a result, comparison of obtained signals from
different codes would be expedited.
Second, the correcting terms will have generically non-trivial angular
dependence. Therefore, the decomposition on different multipolar moments
would be affected. In the case of non-spinning equal mass black holes
symmetry considerations indicate this might not be a significant issue. On
the other hand,
for different masses and/or spinning compact objects the contribution should be
non-trivial, especially in light of significant kicks observed which indicate
a strong direction dependent of the radiated waves.\\

It is important to stress here that the corrections are case-by-case dependent
as they are both sensitive to the gauge and initial data employed.
Consequently it is difficult to assess the role the corrections indicated
here might have. Nevertheless, consistency with the calculation formalism
and a simplified frame of comparison among different implementations are already 
key reasons for the relevance of the discussion. \\

As a side comment we want to stress we have employed a convention
based on a $(-2)$ signature following~\cite{Pirani64,Geroch73} which is
the standard signature employed in studies of asymptotically flat spacetimes. \\

Last, short of considering the corrections discussed in this work, as discussed
in \cite{baumgarteetal}, one could estimate whether these
effects may play a role by evaluating the norms $||{\cal R} - 2||$,
$||g_{ur} - \kappa||$ (with $\kappa=1$ or $\sqrt{2}$ depending on
the boost freedom adopted for $\ell^a$) and $||g_{uA}||$. If these norms
are at the order of the truncation error in a simulation, then
the correcting factors would certainly not be essential. We will examine
these issues for different binary systems in a forthcoming work\cite{testcorrection}.


\section{Acknowledgments}
We would like to thank D. Garfinkle, C. Palenzuela, F. Pretorius, J. Pullin
M. Tiglio and J. Winicour for comments and discussions. This
work was supported in part by grants from NSF: PHY-0326311 and
PHY-0554793 to Louisiana State University and
ANPCyT, CONICET and SeCyT-UNC.
L.L. wishes to thank the University of Cordoba for hospitality
where parts of this work were completed.


\end{document}